%% file: main.tex
\documentclass{svproc}
\usepackage{url}

\usepackage{cite}
\usepackage{amsmath,amssymb,amsfonts}
\usepackage{graphicx}
\usepackage{textcomp}
\usepackage{xcolor}
\usepackage [english]{babel}
\usepackage [autostyle, english = american]{csquotes}
\MakeOuterQuote{"}
\usepackage{float}
\usepackage{graphics} % for pdf, bitmapped graphics files
\usepackage{epsfig} 
\usepackage{subfigure}
\usepackage{tabularx}
\usepackage{multicol}
\usepackage{multirow}
\usepackage{tabularx}
\usepackage{pdflscape}
\usepackage{bbding}
\usepackage{wasysym}
\usepackage{pifont}% http://ctan.org/pkg/pifont
\usepackage{afterpage}
\usepackage{capt-of}
\usepackage{rotating}
\usepackage{lscape}
\usepackage{hyperref}
\usepackage[capitalise]{cleveref}
\usepackage{pgfplots}
\pgfplotsset{width=8cm,compat=1.9}
\graphicspath{ {figs/} }
\def\BibTeX{{\rm B\kern-.05em{\sc i\kern-.025em b}\kern-.08em
    T\kern-.1667em\lower.7ex\hbox{E}\kern-.125emX}}
    
\makeatletter
\patchcmd{\@maketitle}
  {\addvspace{0.5\baselineskip}\egroup}
  {\addvspace{-1\baselineskip}\egroup}
  {}
  {}
\makeatother

\begin{document}
\mainmatter

\title{iGateLink: A Gateway Library for Linking IoT, Edge, Fog and Cloud Computing Environments}
\titlerunning{iGateLink}  % abbreviated title (for running head)
\author{Riccardo Mancini\inst{1,2} \and
Shreshth Tuli\inst{1,3} \and
Tommaso Cucinotta\inst{2} \and
Rajkumar Buyya\inst{1}}
\authorrunning{Riccardo Mancini et al.} % abbreviated author list (for running head)
\tocauthor{Riccardo Mancini, Shreshth Tuli, Tommaso Cucinotta and Rajkumar Buyya}
\institute{Cloud Computing and Distributed Systems (CLOUDS) Lab, School of Computing and Information System, The University of Melbourne, Australia
\and 
Scuola Superiore Sant'Anna, Pisa, Italy
\and
Department of Computer Science and Engineering, Indian Institute of Technology, Delhi, India}

% \DeclareRobustCommand*{\IEEEauthorrefmark}[1]{%
%   \raisebox{0pt}[0pt][0pt]{\textsuperscript{\footnotesize #1}}%
% }

% \author{
% \IEEEauthorblockN{Riccardo Mancini\IEEEauthorrefmark{1,}\IEEEauthorrefmark{2},
% Shreshth Tuli\IEEEauthorrefmark{1,}\IEEEauthorrefmark{3},
% Tommaso Cucinotta\IEEEauthorrefmark{2} and
% Rajkumar Buyya\IEEEauthorrefmark{1}}
% \IEEEauthorblockA{\IEEEauthorrefmark{1}Cloud Computing and Distributed Systems (CLOUDS) Lab, School of Computing and Information Systems\\
% The University of Melbourne, Australia}
% \IEEEauthorblockA{\IEEEauthorrefmark{2}Scuola Superiore Sant'Anna, Pisa, Italy}
% \IEEEauthorblockA{\IEEEauthorrefmark{3}Department of Computer Science and Engineering, Indian Institute of Technology Delhi, India}
% }

\maketitle

\begin{abstract}
  In recent years, the Internet of Things (IoT) has been growing in popularity, along with the increasingly important role played by IoT gateways, mediating the interactions among a plethora of heterogeneous IoT devices and cloud services. In this paper, we present \emph{iGateLink}, an open-source Android library easing the development of Android applications acting as a gateway between IoT devices and Edge/Fog/Cloud Computing environments. Thanks to its pluggable design, modules providing connectivity with a number of devices acting as data sources or Fog/Cloud frameworks can be easily reused for different applications.
%%and, furthermore, the library already provides out-of-the-box implementations for some common data sources and some Fog/Cloud frameworks from literature. 
% The core of the library is written in plain Java for easy portability to different applications, while some Android-specific modules take care of all subtleties regarding Android application life-cycle and thread management. 
  %%The proposed library has been tested through the development of two case-study applications replicating previous works: the first healthcare application connects to a \emph{Bluetooth LE} oximeter to collect and analyze its data; the second, takes a photo and sends it to the Fog/Cloud for object detection. These case studies show how easily the library can adapt to different scenarios and how development is faster when using the library, compared to writing the applications with conventional methods.
Using \emph{iGateLink} in two case-studies replicating previous works in the healthcare and image processing domains, the library proved to be effective in adapting to different scenarios and speeding up development of gateway applications, as compared to the use of conventional methods.
\end{abstract}


%%%%%%%%%%%%%%%%%%%%%%%%%%%%%%%%%%%%%%%%%%%%%%%%%%%%%%%%%%%%%%%%%%%%%%%%%%%%%%%%
\input{introduction}
\input{relwork} % move to bottom ? -> Only if you need to use some terminology that you describe in sections 3 or 4 in this section (but try to avoid as related work should not have complex terms)
\input{model}
\input{implementation}
\input{casestudies}
\input{conclusions}
%%%%%%%%%%%%%%%%%%%%%%%%%%%%%%%%%%%%%%%%%%%%%%%%%%%%%%%%%%%%%%%%%%%%%%%%%%%%%%%%

\bibliographystyle{splncs03_unsrt}

\bibliography{references}

\end{document}

%% file: introduction.tex
\section{Introduction}
\label{sec:introduction}

Recently, the Internet of Things (IoT) has gained significant popularity among both industry and academia, constituting a foundational technology for creating novel computing environments like smart cities and smart healthcare applications, which pose higher and higher requirements on the capabilities of modern computating infrastructures~\cite{yi2015survey, gill2019transformative}. Cloud Computing allowed for offloading complex and heavy-weight computations, including big-data processing pipelines, to remote data centers~\cite{yi2015survey}. However, the exponential growth of connected IoT devices and the forecast in the produced data volumes for the upcoming years~\cite{ericsson2018-edge} pushed industry and academia to look into optimized solutions, where virtual machines are dropped in favour of more light-weight containers~\cite{Cucinotta19-EDGE} and, more importantly, \emph{decentralized solutions} are employed, where computing happens at the \emph{edge} of the network, giving rise to the Fog computing paradigm. This leads to reduced latency, deployment costs and improved robustness~\cite{bonomi2012fog}, so in recent years many fog/cloud frameworks have been proposed leveraging computing resources both at the edge of the network and in cloud data centers \cite{fogbus, edgelens}. 

In nowadays IoT and fog computing frameworks, a crucial component is the \emph{gateway} device which enables communications of users, sensors and actuators with edge devices and cloud resources~\cite{aazam2014fog}. Gateway devices could be small embedded computers, smart routers or even smart-phones. In IoT the number of sensors and actuators have increased tremendously in the last few years~\cite{singh2014survey, lee2015internet}, including advanced gateway devices that emerged in recent fog environments~\cite{whiteaker2012expanding}. Even though fog computing frameworks greatly simplify engineering gateway functionality, they do not focus on making generic gateway interfaces for seamless integration with diverse applications, user needs and computation models.
%% As fog applications increase and more users embrace fog systems, there is an increasing requirement of a generic fog gateway interface which is capable of handling diverse environments.

\begin{figure}[htb]
    \centering
    \includegraphics[width=0.5\textwidth]{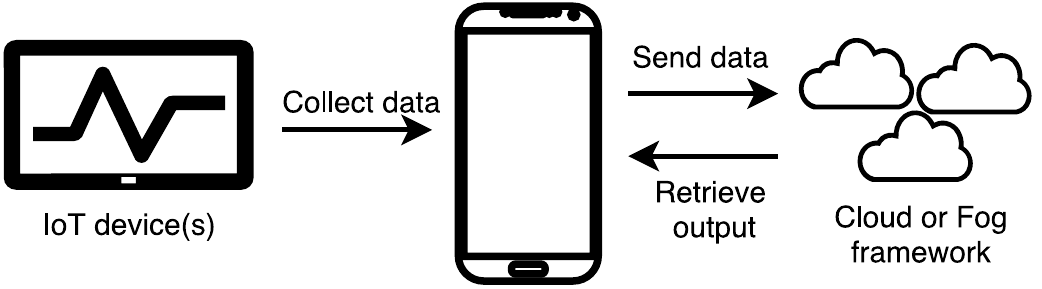}
    \caption{Example scenario in which one or more IoT device
        communicate to the Cloud/Fog through a gateway device.}
    \label{fig:scenario}
\end{figure}

% For a typical fog computing scenario as shown in \Cref{fig:scenario}, various IoT devices communicate with fog or cloud computing nodes in an end-to-end integrated environment through a gateway node (generally a smartphone) in the framework \cite{fogbus, lee2016gateway}. The gateway collects the data coming from the IoT devices and 
% sends it to the fog or cloud nodes in order to be processed. Finally, the result of the execution is retrieved by the gateway device and shown to the user.
% While the scenario itself is common, the technologies used in the IoT devices and Fog/Cloud could be very different, making the  software running on the gateway device specific to the application. That is the reason why the \emph{iGateLink} library was created as a part of this work. 

\paragraph{Contributions.} 

This paper proposes \emph{iGateLink}, a modular fog-cloud gateway library
easing the development of applications running on Android-based IoT gateway devices. It provides common core functionality, so as to let developers focus on the application-specific code, for example implementing communications with specific sensors or protocols to exchange data with specific
fog or cloud systems. \emph{iGateLink} is specific to the mentioned IoT-Fog use-case but generic enough in order to allow simple extensions to be used in different IoT-Fog integrated environments as shown in \Cref{fig:scenario}. It is also easy-to-use through a simple API and supports integration of different frameworks within the same application, including the possibility to run the required execution locally.

\emph{iGateLink} has been applied to two use-cases, one dealing with a oximeter-based healthcare application for heart care patients, the other one for low response time object recognition in camera images. The presented framework proved to be effective for reducing development complexity and time, when comparing with existing practises in coding IoT gateway applications.

% The rest of the paper is organized as follows. \Cref{sec:relwork} presents the existing and related fog or cloud gateway applications and other mainstream applications for generic framework deployment. The model of the proposed library is described in \Cref{sec:model} with implementation details in \Cref{sec:implementation}. In \Cref{sec:casestudies} we demonstrate the efficacy of the library on two case studies. Finally, in \Cref{sec:conclusions} we provide conclusions and future directions.

% estimate 1 page (with title et al.)

%% file: relwork.tex
\section{Related Work}
\label{sec:relwork}

Due to the vast heterogeneity in the Internet of Things, the importance of the IoT 
gateway in enabling the IoT has been acknowledged by several works~\cite{zhu2010iot, datta2014iot, chen2011brief}.
Some authors propose IoT gateways to act only as routers, e.g. incapsulating raw data coming 
from \emph{Bluetooth} devices in \emph{IPv6} packets to be sent to the Cloud~\cite{zachariah2015internet}.
However, in order to enable more complex scenarios, offload
computations and/or save network bandwidth, IoT gateways need to become 
``smart'' by pre-processing incoming data from the sensors~\cite{saxena2017efficient}.
In this context, the use of smartphones as IoT gateways has been proposed~\cite{golchay2011, aloi2016mobile}, however those works do not take into
consideration the most recent Fog and Edge Computing paradigms.

Regarding efforts to integrate the IoT with Fog and Edge Computing, \emph{Aazam et al.}~\cite{aazam2014fog} proposed a Smart Gateway based communication that utilizes data trimming and pre-processing, along with Fog computing in order to help lessen the burden on the cloud.
Furthermore, \emph{Gia et al.}~\cite{gia2015fog} developed a Smart IoT Gateway for a Smart Healthcare use-case.
Finally, \emph{Tuli et al.}~\cite{fogbus} developed \emph{FogBus} an integration framework for IoT and Fog/Cloud. However, their work does not provide a generic application able to integrate IoT
and Fog/Cloud frameworks, building their application from the ground-up, tailored to their
specific use-case.

Finally, it is worthwhile to note that none of the previously mentioned works focuses on the
design and implementation of the software that is required to run in the IoT gateway,
especially in the case of an Android device, in order to make it generic and adaptable 
to many different scenarios, as this paper does.

% I think this section needs to be more comprehensive discussing more works and arguing why iGateLink is required preferably with a table showing the features of such gateway application how current works lack various features which this work provides
% The reader should get convinced that with iGateLink, fog computing would really benefit

% estimate 0.5 pages

%% file: model.tex
\section{System Model and Architecture}
\label{sec:model}

% Structure:
% a. "Problem"
%    - why a generic library? which benefits does it bring with? 
%    - fig?
% b. requirements given the problem
%    1. generic/modular/extendable
%    2. specific to the fog/cloud use-case
%    3. support for multiple frameworks and local execution => Chooser
%    4. easy to use ?
% c. resulting design decisions
%    - choice of the observer/pub-sub pattern
%    - chooser mechanism (why? what capabilities does it enable?)

% \begin{figure}
%     \centering
%     \includegraphics[width=0.9\columnwidth]{figs/fgl_library_overview.pdf}
%     \caption{Schematic overview of the requirements and features of the
%     library.}
%     \label{fig:overview}
% \end{figure}

The principles discussed in \Cref{sec:introduction} have been 
addressed by using a modular 
design, with a generic core upon which different modules can be 
loaded. 

%%In order to allow this modular design,
A variation of the publish/subscribe paradigm~\cite{eugster2003many}
has been used, where the components collecting data from 
sensors and the ones sending it to the Fog/Cloud are the publishers 
and are called \emph{Providers}, 
while the subscribers are the auxiliary components that manage 
the execution of the publishers or UI components that show incoming 
data to the user.
The publish/subscribe paradigm is realized through an intermediate 
component that stores the incoming data from the \emph{Providers}, 
called \emph{Store}, and notifies the subscribers, called \emph{Triggers},
using the \emph{observer} design pattern. A \emph{Store} can also be 
thought as the \emph{topic} to which the \emph{Provider} \emph{publishes}
data, while the \emph{Triggers} are its \emph{subscribers}.

In the considered scenario, a \emph{Provider} may be started either 
as a result of a user interaction (e.g. a button click) or as a 
consequence of external event (e.g. incoming Bluetooth data). 
It can also be started with some input data or with no data at all. 
The proposed model does not assume any of the aforementioned cases, employing
a generic design, able to adapt to many different scenarios.
While the input of a \emph{Provider} can vary, the result is always data that
needs to be stored in a \emph{Store}.
Whenever a \emph{Provider} stores new data to a \emph{Store}, 
all \emph{Triggers} associated with the \emph{Store} are 
executed.  
The most common use of a \emph{Trigger} is to start 
another \emph{Provider} but it could also be used, for example, to 
update the UI.

These design choices enabled: 1) modularity since \emph{Providers} and \emph{Triggers}
can be independently and easily plugged and unplugged; 2) flexibility  
since this model enables even more complicated use-cases than
the one mentioned above; 3) ease of use thanks to code reusability.

Furthermore, several \emph{Providers} providing the same data 
(i.e. publishing to the same \emph{Store}) can be active 
simultaneously, for example \emph{Providers} for different Fog/Cloud frameworks and/or local 
execution. In this case, it is useful to define a new component, 
the \emph{Chooser}, whose function is to select a specific
\emph{Provider} among a list of equivalent ones in order to produce data.
By doing so, it is possible to, for example, use another \emph{Provider} 
when one is busy or to fallback to another \emph{Provider} if one fails.

% You should describe model in more details independent of implementation specific stuff
% like what various modules are (in abstract sense) and how they interact and provide the claimed generic nature, etc.
% estimate 1 pages

%% file: implementation.tex
\subsection{Implementation Details}
\label{sec:implementation}

% Structure:
% a. overview
%    - "onion structure" (fig?) 
% b. core components
%    - basic class diagram and description of the components 
% c. implementation
%    - android-specific implementation of the provider (asynctask)
%    - android foreground service wrapping
% d. extension components ?
%    - bluetooth le ?
%    - camera ?

\begin{figure}
    \minipage{0.4\textwidth}
        \includegraphics[width=\linewidth]{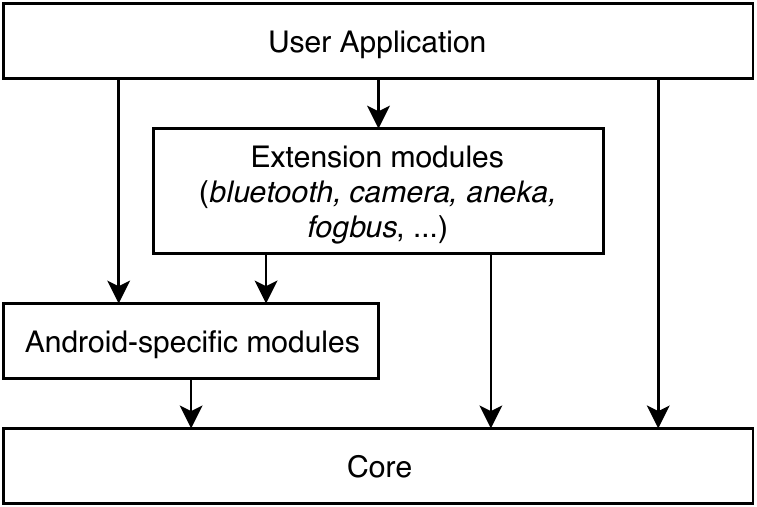}
        \caption{High-level overview of the library software components.}
        \label{fig:stack}
    \endminipage\hfill
    \minipage{0.55\textwidth}
        \includegraphics[width=\linewidth]{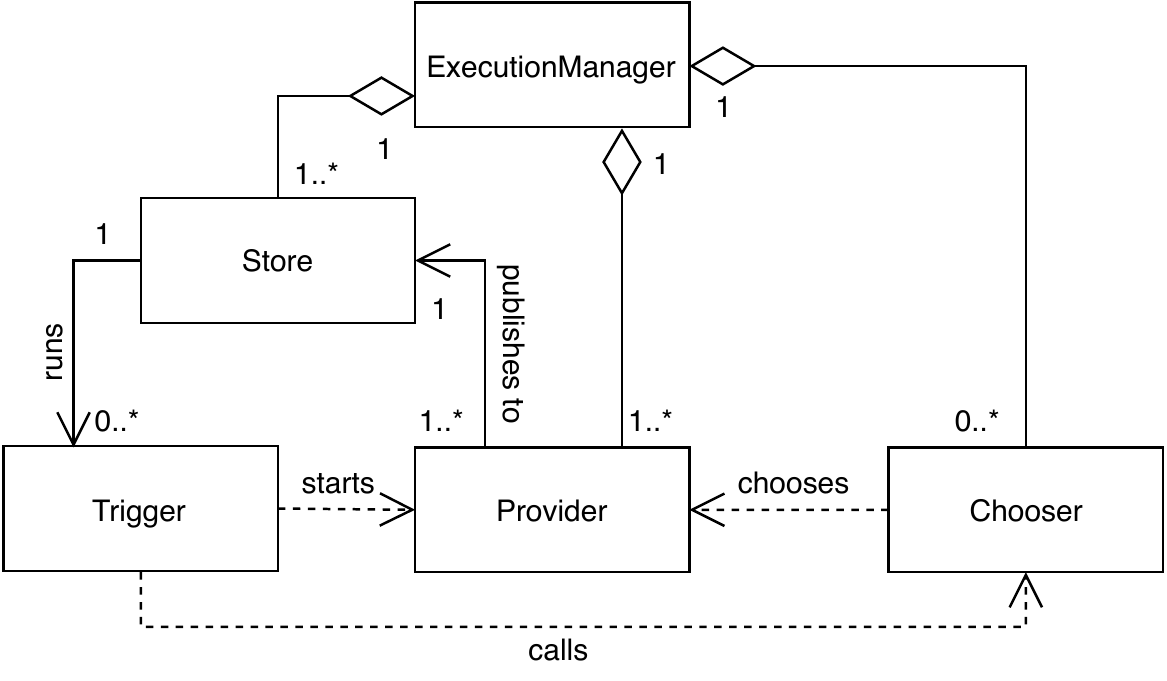}
        \caption{Simplified UML diagram of the core components.}
        \label{fig:core}
    \endminipage
\end{figure}
% bluetooth, fogbus, edgelens are proper nouns should have first letter caps
% they're the name of the modules, which in Java are always lower case. You can either italicise them then
% I removed edgelens because there's no module with that name (it's just 
% implemented as a class).

Based on the model described above, we have implemented 
the \textit{iGateLink} library for Android devices.
The library is open-source and available at: \url{https://github.com/Cloudslab/iGateLink}.
From a high-level point of view (\Cref{fig:stack}), the library is composed 
of a platform-independent core written in Java, an Android-specific module which extends 
the core to be efficiently used in Android
and several extension modules that provide complimentary 
functionalities.

\paragraph{Core.}
\label{subsec:core}
The core of the library is composed by the following classes: 
\emph{ExecutionManager}, \emph{Data}, \emph{Store}, 
\emph{Provider}, \emph{Chooser}
and \emph{Trigger}. Refer to \Cref{fig:core} 
for a conceptual overview of their interactions.
The \emph{ExecutionManager} class
coordinates the other components and provides an API for managing 
them.
The \emph{Data} class
is not properly a component but is the base class that every data 
inside this library must extend. It is characterized by an 
\texttt{id} and a \texttt{request\_id}: the former must be 
unique between data of the same \emph{Store}; the latter is 
useful for tracking data belonging to the same request.
The \emph{Store} class
stores \emph{Data} elements and provides two operations: 
\texttt{retrieve} for retrieving previously stored \emph{Data} 
and \texttt{store} for storing new data. 
Every \emph{Store} is uniquely identified by a \texttt{key}. 
There can be multiple \emph{Stores} for the same \emph{Data} 
type. Furthermore, a \emph{Store} can have one or more \emph{Triggers}
associated to it that is called whenever new data is stored.
For example, a common use for a \emph{Trigger} is
starting a \emph{Provider} with the recently stored \emph{Data} 
from another \emph{Provider}.
The \emph{Provider} class
takes some \emph{Data} in input and produces some other 
\emph{Data} in output. 
Every \emph{Provider} is uniquely identified by a \texttt{key}. 
There can be multiple \emph{Providers} for the same 
\emph{Store} but a \emph{Provider} can only use one 
\emph{Store} at a time. A \emph{Provider} can be started by calling its 
\texttt{execute} method either through \emph{ExecutionManager}'s \texttt{runProvider} 
or \texttt{produceData}. 
In the latter case, the user specifies only which \emph{Data} it 
wants to be produced (by means of a \emph{Store} key) and a 
suitable \emph{Provider} will be executed. 
In case there are two or more \emph{Providers} for the same 
\emph{Store}, a \emph{Chooser} will be used.

\paragraph{Android-specific modules.}
\label{subsec:android}
% Just a general overview, avoid details on implementation -> next section
% - asyncprovider
% - foreground service

When developing such an application
on Android, a common problem is the one to use \emph{worker threads} for time-consuming tasks
that need to execute in the background still letting the Android runtime to kill the app as needed.
This is addressed by providing the 
\emph{AsyncProvider} class, which makes use of the
\emph{AsyncTask} class\footnote{it provides a simple API for 
scheduling a new task and executing a custom function on the
main thread before and after the execution}. 
% Doing so also
% removes the need of any synchronization mechanism in the 
% \emph{Store} since all \texttt{store} and \texttt{retrieve} operations
% would be done on the main thread. 
%
Furthermore, the \emph{ExecutionManager} is hosted
within an Android \emph{Service},
which, if configured correctly as a foreground service, prevents the
Android runtime from killing it.

%% file: casestudies.tex
\section{Case studies}
\label{sec:casestudies}
% description of the implementations with focus to 
% the different protocols and how problems were solved:
% - fogbus
% - edgelens
% - aneka
% Emphasis on how this library provided following benefits:
% - Engineering simplicity in application deployment
% - Enabling users to integrate diverse applications
% - Robust enough to be included in different frameworks
% Consider adding some comparative analysis among different works fogus, edglens, aneka and iGateLink clearly describing the contrast and need for this generic library for each case.

In order to test the library and demonstrate its applicability to real applications,
two case studies have been developed. They both reproduce 
existing works, namely FogBus~\cite{fogbus} and EdgeLens~\cite{edgelens}.
The first application
%(\Cref{subsec:btapp})
connects to a \emph{bluetooth} 
oximeter to collect and analyze its data.
The second application
%(\Cref{subsec:camapp})
takes a photo and sends it to the Fog or 
Cloud for object detection.

In the original works both applications have been developed using 
\emph{MIT App Inventor}, which eases and speeds up the 
development but provides only a very simple API that cannot 
be adapted to more complicated use cases. Furthermore, every 
application needs to be built from the ground up even though
many components may be in common, especially when using the same
input method or framework.
By developing the applications using \emph{iGateLink},
both modularity and fast development can be achieved.

% The structure of both applications is very simple: there is a
% screen where the user can set the application settings, a bottom 
% navigation menu and
% one or more screens specific to the application. 
% The user first inserts the correct settings in the settings screen 
% (see \Cref{fig:oximeter_app}) and then enables the service to 
% start the \emph{ExecutionManager}. Once
% the service is running, the user can interact with the other 
% screens.

% Regarding the implementation of the applications in Android, both 
% applications contain only one \emph{Activity}, with the different
% screens being different \emph{Fragments},
% the configuration screen is a \emph{PreferenceFragment} and 
% navigation between screens is implemented using the \emph{Android 
% Jetpack} \emph{Navigation} component.

% In the following sections, a brief overview of the developed applications
% is provided.

\begin{figure}
    \minipage{0.475\textwidth}
        \includegraphics[width=0.45\linewidth]{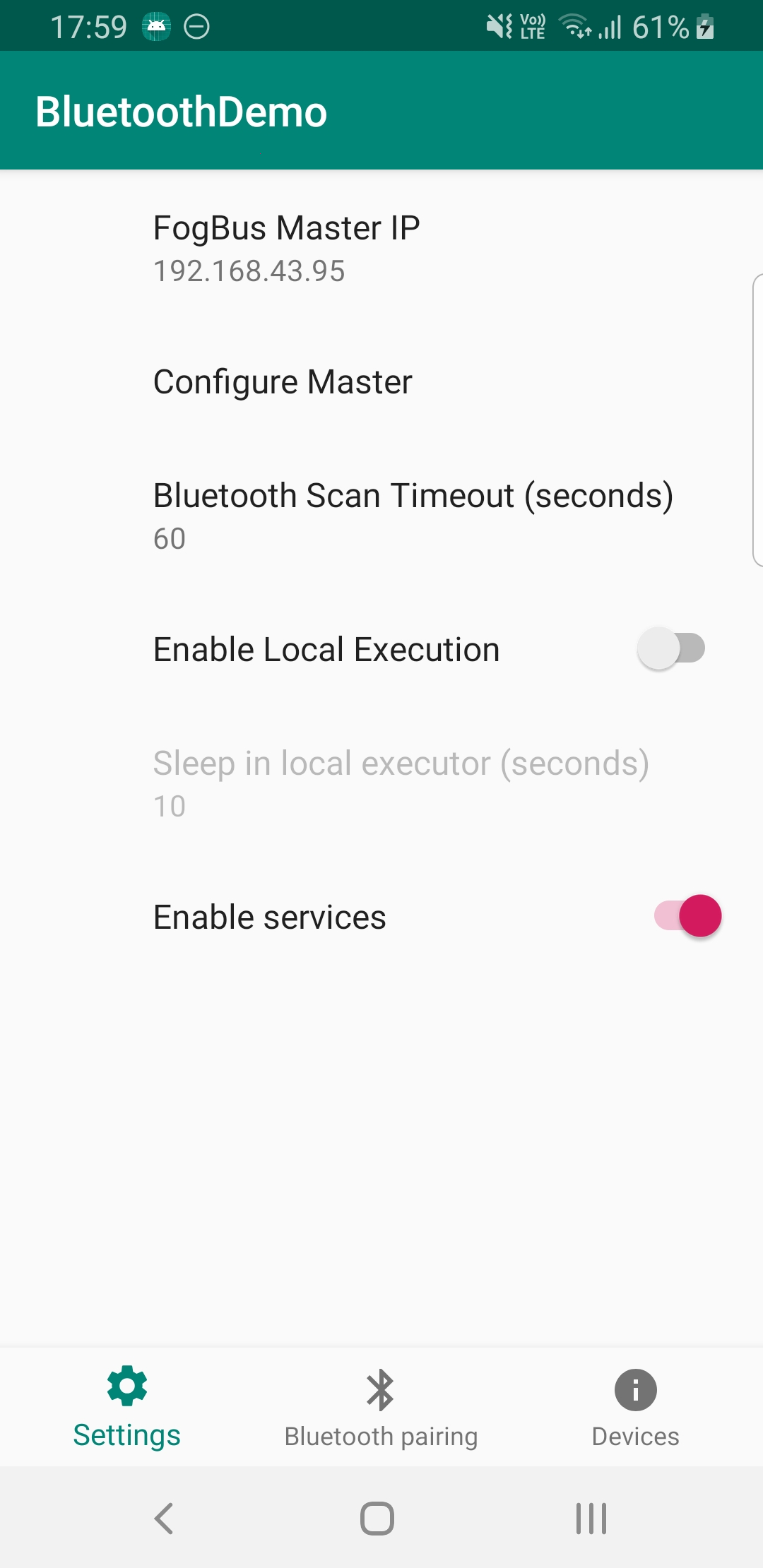}
        \includegraphics[width=0.45\linewidth]{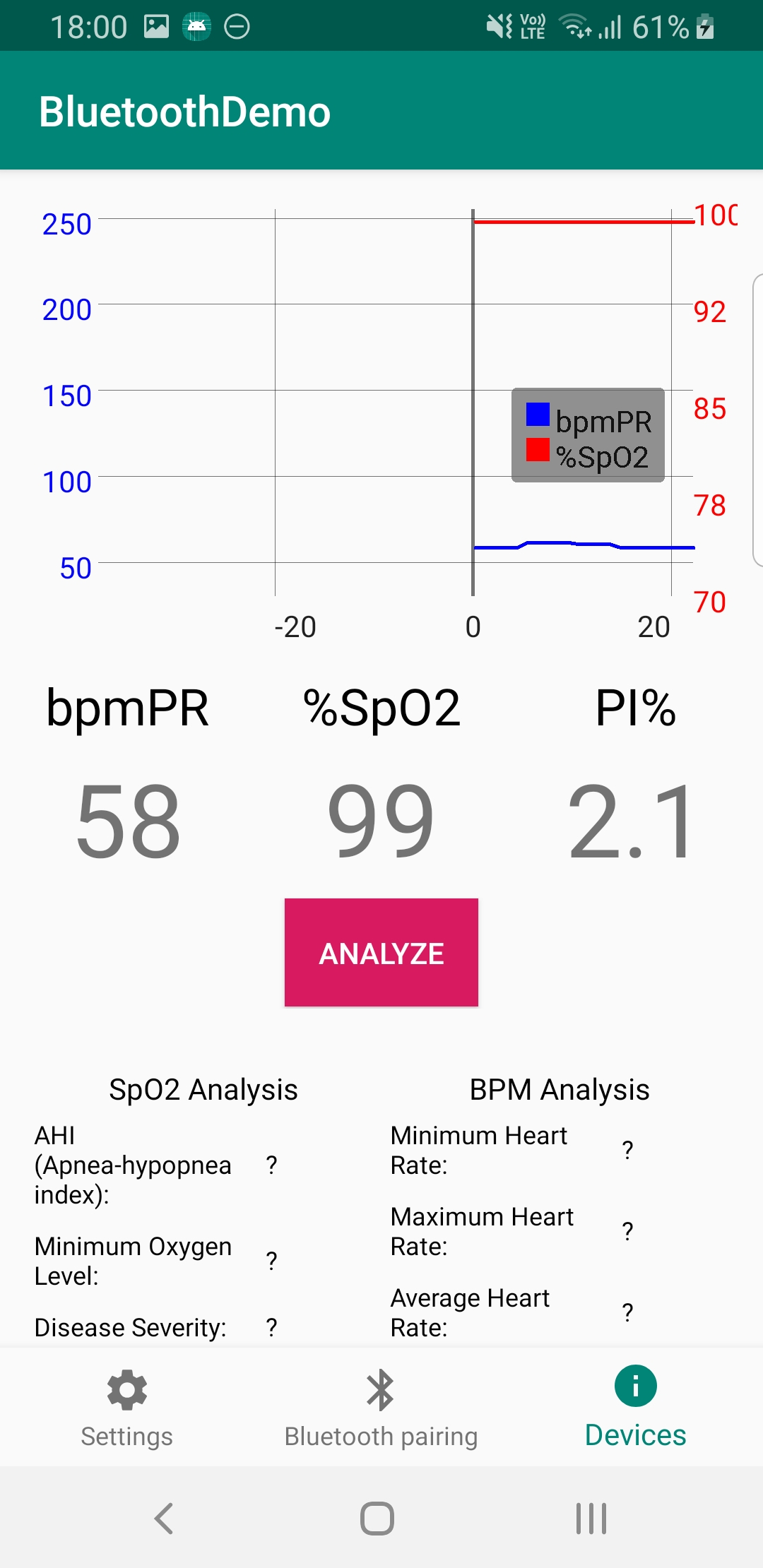}
        \caption{Configuration screen (left) and live data and analysis results screen (right) 
        from the oximeter demo app.}
        \label{fig:oximeter_app}
    \endminipage\hfill
    \minipage{0.475\textwidth}
        \includegraphics[width=0.45\columnwidth]{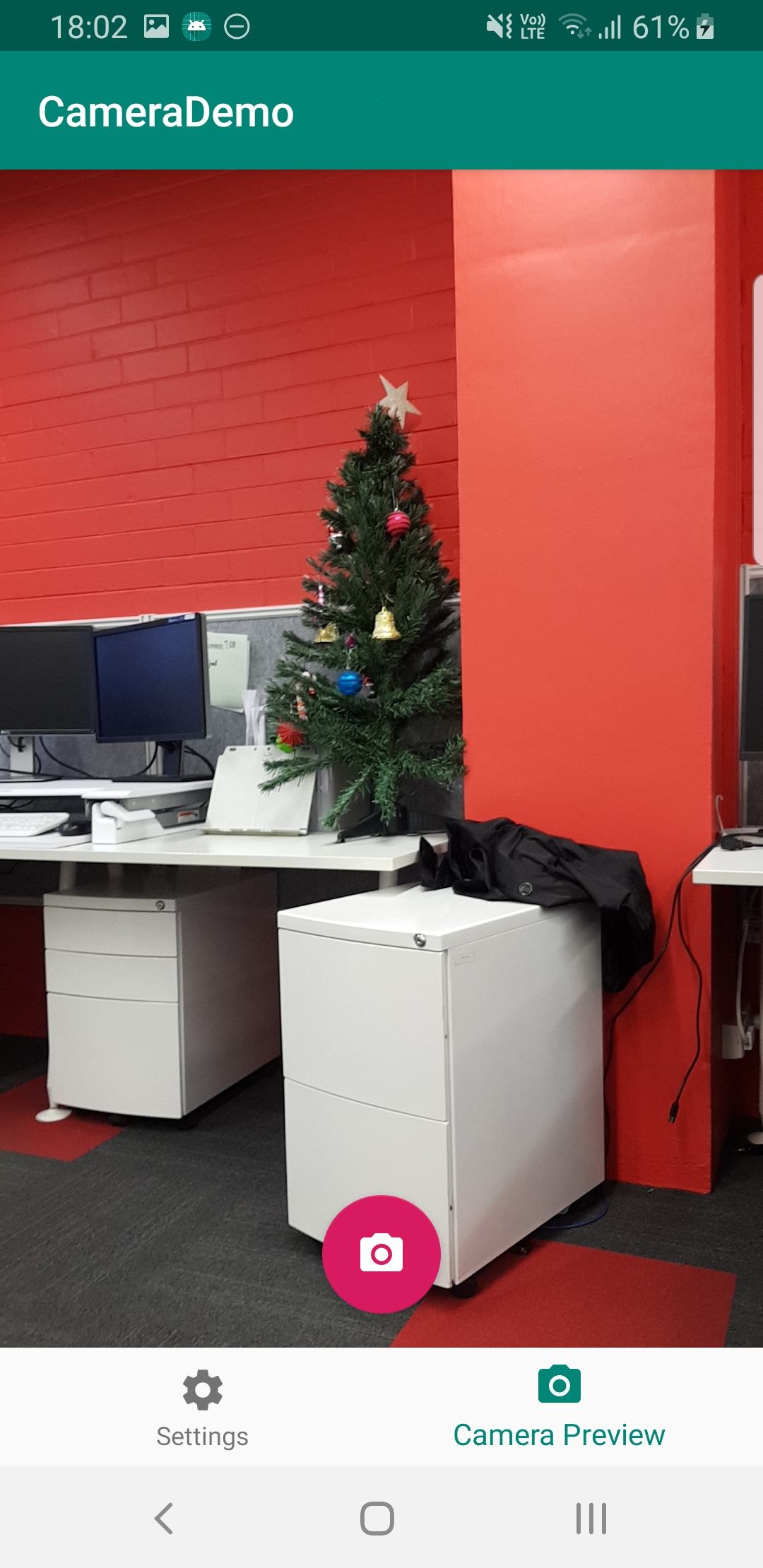}
        \includegraphics[width=0.45\columnwidth]{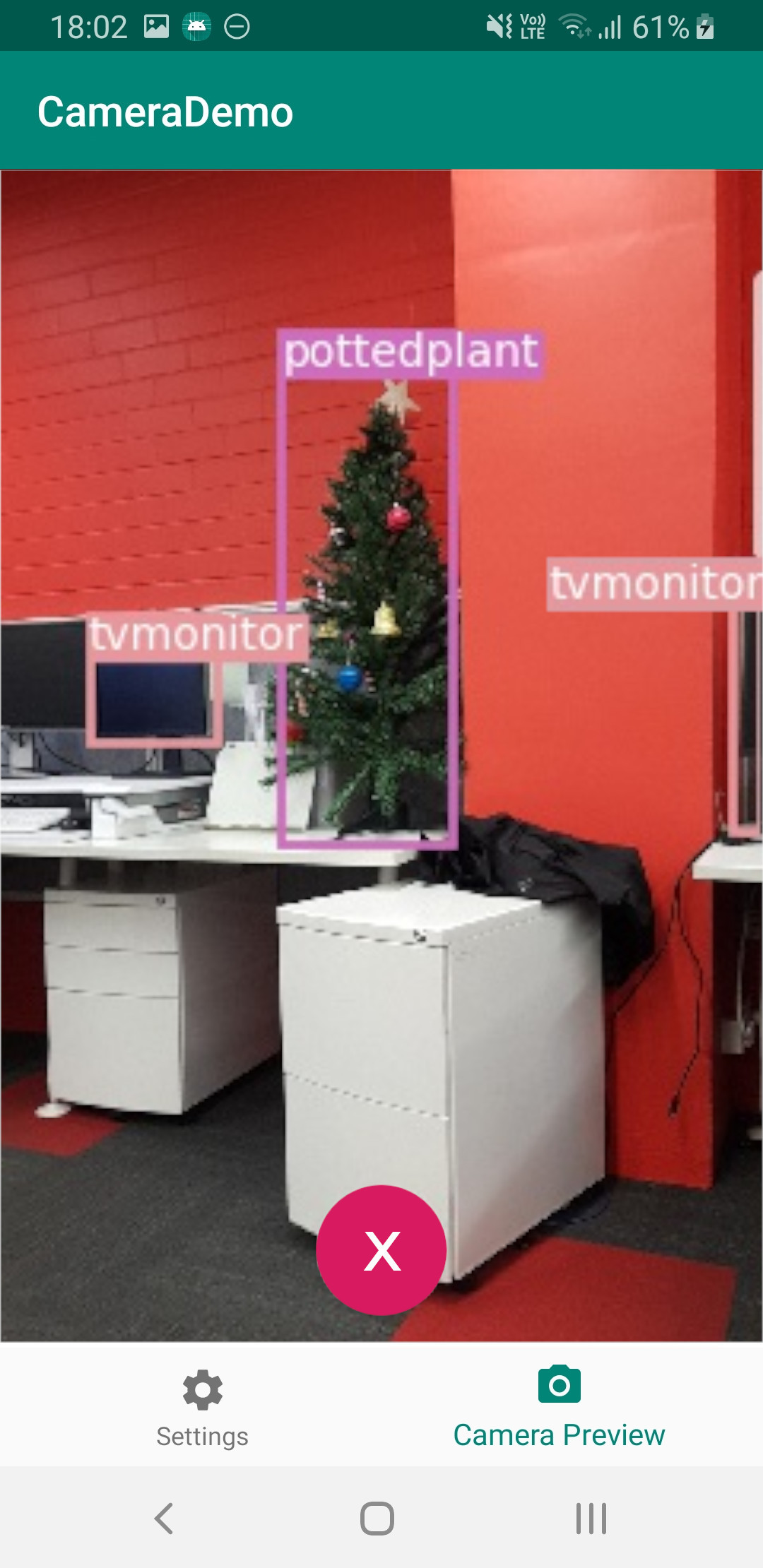}
        \caption{Preview screen (left) and result screen (right) from the object detection demo app.}
        \label{fig:camera_app}
    \endminipage
\end{figure}

\paragraph{Oximeter based healthcare application}
\label{subsec:btapp}

% Fogbus
% -> single http request, master forwards request to worker or cloud
% Bluetooth -> data flow is async
% local execution
The \emph{bluetoothdemo} application can be used to detect hypopnea in a patient 
by collecting data from an oximeter. The application consists of four screens:
1) the configuration screen (left picture in \Cref{fig:oximeter_app});
2) the Bluetooth device pairing screen;
3) the Bluetooth device selection screen through which the user chooses the device whose data to show;
4) the data and analysis result screen (right picture in \Cref{fig:oximeter_app}).

Data is collected in real-time from the oximeter using \emph{Bluetooth LE}. In 
order to do that, the \emph{bluetooth} module has been developed which provides
an implementation of the \emph{Provider} which registers to the \emph{GATT} characteristics
in order to receive push notifications from the \emph{Bluetooth} device. Raw data received
from the oximeter is then converted to a \emph{Data} object and stored.

The user can then upload the data to \emph{Fogbus}~\cite{fogbus}
by tapping the ``Analyze'' button in order to get the analysis results. 
Data is sent using an HTTP request to the \emph{Fogbus} master node which forwards 
the request to a worker node (or the cloud) and then returns the result to the application. 
This simplifies the development of the application, since only one HTTP request is required.

\paragraph{Object detection application}
\label{subsec:camapp}

% Edgelens -> arbiter, upload, exec, output
% Camera -> provider is started by user
% aneka -> login and application creation, ftp upload, job submission, ftp download
The \emph{camerademo} application can be used to take a photo and run an object
detection algorithm on it, namely \emph{Yolo}~\cite{redmon2016you}. The user interface is very
simple, with just a screen showing the camera preview and a button (\Cref{fig:camera_app}). 
When the button is
clicked, the photo is taken and sent to the Fog/Cloud for object detection. When the execution is
terminated, the resulting image is downloaded and shown to the user.

In order to integrate Android camera APIs with \emph{iGateLink}, the \emph{camera} module
has been developed, which provides an implementation of a \emph{Provider}, 
\emph{CameraProvider}, that takes a photo when its \emph{execute} method is called. When 
the photo is stored, two providers are executed: \emph{BitmapProvider}, that converts the photo
from an array of bytes to a \emph{Bitmap} object so that it can be plug in a \emph{ImageView} and 
shown to the user, and one of \emph{EdgeLensProvider} or \emph{AnekaProvider},
that execute the object detection on either \emph{EdgeLens}~\cite{edgelens} or Aneka~\cite{aneka}.
The result of the object detection is, again, an array of bytes so it goes through another
\emph{BitmapProvider} before being shown to the user.

The \emph{EdgeLensProvider} is responsible for uploading the image to the \emph{EdgeLens} framework
and downloading the result once the execution is completed. \emph{EdgeLens} has a similar 
architecture to \emph{Fogbus} but, differently from it, communication between client and worker
is direct, instead of being proxied by the master. The usual \emph{EdgeLens} workflow is:
1) query the master to get the designated worker;
2) upload the image to the worker;
3) start execution;
4) download the result when the execution is terminated.

The \emph{AnekaProvider} is responsible for execution in 
\emph{Aneka}~\cite{aneka} through its REST APIs for task submission. 
% The workflow is similar to the one of \emph{EdgeLens},
% in the sense that three steps are necessary (upload, execute, download),
% and to the one of \emph{FogBus}, since communication happens only with the master node.
The image is first uploaded to a FTP server (which could be hosted by the master node itself), then 
the object detection task is submitted to \emph{Aneka}, whose master node chooses a worker node to submit the request to. 
The worker downloads the image from the FTP server, runs the object detection on it and uploads the
result image back to the FTP server. 
In the meanwhile, the client repeatedly polls the master node in a loop, waiting for the submitted task to complete. When it does, the client finally downloads the result which is displayed to the user.

% Pls add a summary of all points in different case studies in the end of this section

% estimate 2 pages

%% file: conclusions.tex
\section{Conclusions and Future Work}
\label{sec:conclusions}

In this paper we presented a new Android library, \emph{iGateLink}, which
enables developers to easily write new Android applications linking 
IoT, Edge, Fog and Cloud Computing environments, as it comes with a set
of highly reusable modules for common IoT scenarios.
%%It has been build to be 
%%generic and modular. Modules can be easily reused in many application 
%%scenarios, enhancing code reusability and development time.

We demonstrated, by means of two use-cases, that the library can 
adapt to different applications.
%%, being built to be generic and modular.
The library is easy to use, increasing the engineering simplicity of application deployments, making such systems robust and easy to maintain. 

As part of future work, more modules could be added to the library to cover the most
common use-cases in order to make the development of a new application using the 
library easier. For instance, the current version of the library does not include any module providing a 
\emph{Provider} for Bluetooth devices (currently only Bluetooth Low-Energy is supported), built-in sensors (for example accelerometer, gyroscope, magnetometer, 
luminosity), touch events, audio recording. 
Furthermore, more Fog/Cloud frameworks could be integrated within the library, making it like plug-and-play software for end users. 
% In addition, the existing library core, being written in plain Java, 
% could be ported to different devices, for example personal computers 
% or ARM based devices like Raspberry Pi. 

% Write some points later when other sections are complete
% estimate 0.5 page ?

% Pls fill this, I will update once you are done

% \section*{Software Availability}
% We released \textit{iGateLink} as an open source software. The implementation code with experiment scripts and results can be found at the GitHub repository: \url{https://github.com/Cloudslab/iGateLink}.

% Pls transfer the github repo to cloudslab page https://github.com/Cloudslab/ and rename to iGateLink (for transferring a repo, see settings in github page)
% I'll do that after submission

% \section*{Acknowledgements}
% This research work is supported by the Melbourne-Chindia Cloud Computing (MC3) Research Network and Australian Research Council. We would like to thank --- for their valuable comments on improving the quality of presentation. 